\documentclass[pra,aps,showpacs,superscriptaddress,twocolumn]{revtex4-2}
\usepackage{graphicx,amsmath,amssymb, color, braket}
\usepackage{xcolor}
\usepackage{svg}
\usepackage[colorlinks, citecolor={blue!50!black}, urlcolor={blue!50!black}, linkcolor={red!50!black}]{hyperref}

\begin{document}
\title{Microwave‑Controlled Photonic Spin Hall Effect in Atomic System and Microwave Electrometry}
\author{Muhammad Waseem }
\email{mwaseem@pieas.edu.pk}
\affiliation{Quantum Optics Lab, Department of Physics and Applied Mathematics, Pakistan Institute of Engineering and Applied Sciences (PIEAS), Nilore, Islamabad $45650$, Pakistan.}
\affiliation{Center for Mathematical Sciences, PIEAS, Nilore, Islamabad $45650$, Pakistan}
\date{\today }

\begin{abstract} 
The photonic Spin Hall Effect (SHE) causes a polarization-dependent transverse shift of light at an interface.  
In this paper, we theoretically investigate the microwave‑controlled photonic SHE and microwave electrometry in a closed-loop $\Lambda$-type atomic system. 
At zero probe detuning ($\Delta_p = 0$) and relative phase $\phi=0,\pi$, the photonic SHE magnitude remains at the intrinsic finite-beam saturation limit $w_0/2$, whereas the microwave field shifts its angular position linearly with opposite slopes (signs) for the two phases, enabling differential microwave electric-field sensing.
At $\phi=\pi/2$, the photonic SHE magnitude decays exponentially with $\Omega_\mu$, providing weak microwave electric field sensing.
In contrast, at $\Delta_p = \pm \Omega_c$ (control field coupling), the photonic SHE reaches a maximum of half the incident beam waist when $\Omega_{\mu} \geq \Omega_p$ and the phase $\phi$ is optimized.
Interestingly, we observed microwave-controlled switching of photonic SHE by tuning the relative phase $\phi$ at an optimized value of $\Omega_{\mu}$ and $\Omega_{c}$.
Our results may have potential applications in microwave quantum sensing and quantum optical switches based on the photonic SHE.
\end{abstract}
\maketitle
\begin{section}{Introduction}
Atomic coherence effects in quantum optical systems, induced by coherent electromagnetic fields, have attracted considerable attention due to intriguing counterintuitive physics and potentially important applications~\cite{scully1997quantum}. 
Traditionally, such coherence is established by employing optical fields, which can induce constructive and destructive quantum interference between atomic energy levels. 
However, in the past decades, the addition of microwave fields in these quantum optical systems has introduced a new method of control and tunability, leading to novel coherence effects and deepening the understanding of light–matter interaction~\cite{kosachiov_efficient_2000,wilson_perturbing_2005}.
Microwave-field-induced coherence enables dynamic control over the optical properties of atomic media, building on foundational works in electromagnetically induced transparency (EIT)~\cite{scullyPRA,rostovtsev_electromagnetically_2006} and four-wave mixing (FWM) in atomic vapors~\cite{zibrov_four-wave_2002}. 

The approach of microwave coherence control has since been employed to engineer a wide range of quantum optical phenomena. 
For example, it has enabled efficient control for advanced Rydberg-state-based microwave electrometry~\cite{PhysRevApplied.22.024034,zhou_improving_2023,sedlacek_microwave_2012}, magnetometry~\cite{kiehl_accurate_2025}, trap loss spectroscopy~\cite{cao_trap_2025,duverger_metrology_2024}, and even in the design of microwave-enhanced quantum heat engines~\cite{heatPhysRevA}.
The influence of strong microwave-induced nonlinearities has also been investigated for higher-order squeezing in modified EIT systems~\cite{sqzPhysRevA.104.013706}, non-degenerate four-wave mixing processes~\cite{Mallick_2024}, enhanced coherent population trapping for quantum state preparation~\cite{liu_rabi_2021}, and for achieving tunable optical bistability through amplitude and phase modulation of the microwave field~\cite{cheng_optical_2006}.
Additionally, microwave-induced coherence has enabled the realization of a giant self-Kerr nonlinearity~\cite{beena_2022}, the generation of atomic gratings~\cite{Saddique_2021}, and broadband radiation transport in dense media~\cite{barantsev_broadband_2017}. Applications further extend to control of group velocity for slow and fast light propagation~\cite{Luo:10}, Goos-H\"{a}nchen and Imbert-Fedorov shift~\cite{zhang_controlling_2017}, transient absorption and dispersion tailoring~\cite{zeng_amplitude_2013}, and the formation of slow-light solitons~\cite{li_slow-light_2010}. 
Moreover, this technique has enabled precise microwave electric-field measurement using active Raman gain~\cite{Yang:19}, as well as for the generation of structured and vortex light beams in rubidium vapor under EIT conditions~\cite{verma, Mallick:24, sharma}.
Together, these developments highlight the pivotal role of microwave-assisted atomic coherence in advancing quantum optics and light–matter interaction research.

Recently, the atomic coherence effects have been exploited to manipulate the photonic SHE of a light beam reflected or refracted from the interface of atomic media~\cite{abbas_tunable_2025,waseem_gain-assisted_2024,shah_coherent_2025,khan_loss-free_2025}, atomic-ensemble-based cavity optomechanics~\cite{munir_enhanced_2025}, and anisotropic two-dimensional atomic crystals~\cite{zhang_photonic_2022}.
The photonic SHE originates due to the spin-orbit interaction of light and causes a transverse displacement of circularly polarized components upon reflection or refraction at an interface~\cite{zhu_wave-vector-varying_2021,he_high-order_2024,sheng_photonic_2023,liu_photonic_2022}.
The photonic SHE is considered an analog of the electronic SHE~\cite{hirsch1999spin}. 
In photonic SHE, the spin and refractive index gradient of the photons play roles similar to the spin and electric potential of the electrons, respectively~\cite{bliokh_conservation_2006, hosten_observation_2008}.
Since its initial observations~\cite{bliokh_conservation_2006,hosten_observation_2008}, photonic SHE has been studied in a variety of physical platforms. For example, semiconductors~\cite{menard_imaging_2009}, graphene layers~\cite{kort-kamp_photonic_2018,zhou_identifying_2012, SHAH2024107676,chen_precision_2020,abbas_graphene_2025}, surface plasmon resonance systems~\cite{salasnich2012enhancement,tan_enhancing_2016,wan_controlling_2020}, metamaterials~\cite{yin_photonic_2013}, all-dielectric metasurfaces~\cite{kim_reaching_2022,zhen_controlling_2023,Ma:23}, topological insulators \cite{Shah_2022,zhou_photonic_2013}, strained Weyl semimetals \cite{jia2021tunable}, hyperbolic metamaterials \cite{kapitanova2014photonic}, two-dimensional quantum materials \cite{Shah_2022,kort-kamp_topological_2017,xie_photonic_2025}, cavity magnomechanical system~\cite{abbas_magnomechanics_2025}, PT-symmetric non-Hermitian cavity magnomechanics system~\cite{fahad_photonic_2026}, and in a parity-time (PT) symmetric system with balanced gain and loss \cite{zhou_controlling_2019}. Photonic SHE has potential applications in quantum sensing~\cite{zhou_identifying_2012}, precision measurement~\cite{chen_precision_2020}, development of coherent quantum light sources~\cite{benelajla_physical_2021,steindl_cross-polarization-extinction_2023} and contrast imaging~\cite{he_high-order_2024}.

In coherently prepared atomic systems, spin-dependent shifts have primarily been controlled through optical field parameters, such as probe-field detuning and Rabi frequencies, while exploiting EIT or related coherence effects to suppress absorption losses~\cite{waseem_gain-assisted_2024,abbas_tunable_2025,khan_loss-free_2025,shah_coherent_2025,wan_controlling_2020}.
Ref.\cite{waseem_gain-assisted_2024} investigated gain-assisted coherent control of the photonic SHE across the Brewster angle in both anomalous and normal dispersion regimes, where the optical control field enables tunability of the photonic SHE, but the resulting signal remains relatively weak.
A tripod atom–light configuration\cite{abbas_tunable_2025} showed that the photonic SHE is enhanced near the Brewster angle when all fields are resonant, whereas it becomes notably weaker under off-resonant conditions.
A comparative control architecture in CTL, $\Lambda$-type, and $N$-type atomic configurations has been employed to identify which configurations lead to a photonic SHE that does or does not depend on atomic density and control-field strengths at resonance~\cite{shah_coherent_2025}.
Another important study demonstrated the enhancement of the photonic SHE due to the excitation of surface plasmon resonance in a Kretschmann configuration containing an $N$-type coherent atomic medium, where the magnitude and sign of the photonic SHE depend on internal damping and absorption~\cite{wan_controlling_2020}.

As mentioned earlier, microwave coherence control in quantum optics has been employed for high-sensitivity Rydberg-state-based microwave electrometry by directly measuring the absorption or dispersion of the atomic medium~\cite{PhysRevApplied.22.024034,zhou_improving_2023,sedlacek_microwave_2012,Yang:19}.
The photonic SHE, however, provides a complementary spatial and polarization-resolved readout. Moreover, its small transverse displacement can be amplified through weak measurement and detected using standard polarization optics and a position-sensitive detector~\cite{hosten_observation_2008, liu_photonic_2022}.
To the best of our knowledge, we present the first study of microwave control of the photonic SHE and microwave electrometry based on the photonic SHE in an atomic system.
In a closed-loop $\Lambda$-type atomic system, we show that the microwave Rabi frequency $\Omega_\mu$ and the relative phase $\phi$ between the optical and microwave fields reveal distinct features of the photonic SHE compared to previous studies~\cite{PhysRevApplied.22.024034,zhou_improving_2023,sedlacek_microwave_2012,Yang:19}.
First, at $\Delta_p=0$ and $\phi=0,\pi$, the microwave field linearly shifted the angular position of the photonic SHE and the Brewster angle, while the magnitude of the photonic SHE remained constant at the intrinsic finite-beam saturation limit $w_0/2$.
In this regime, the two-phase values produce angular tuning of equal magnitude but opposite sign, enabling differential microwave sensing.
Second, for intermediate phase values, increasing $\Omega_{\mu}$ exponentially decreases the magnitude of the photonic SHE, with a large exponential coefficient that can be exploited for weak microwave-field sensing.
In the off-resonant regime at $\Delta_p = \pm \Omega_c$, where $\Omega_c$ is the control-field Rabi frequency, the photonic SHE can be tuned to its maximum upper limit for $\Omega_{\mu} \geq \Omega_p$ and optimized values of $\phi$, corresponding to a unit refractive index.
Finally, we demonstrate phase-selective switching of the photonic SHE between nearly zero and its maximum upper limit ($w_0/2$) at optimized values of $\Omega_c$ and $\Omega_{\mu}$.

The rest of the paper is organized as follows: Section II presents a detailed theoretical model, while Section III presents the results and analysis.  
Finally, section IV summarizes the conclusions.

\end{section}
\begin{figure}[t]
\includegraphics[width=3.25 in]{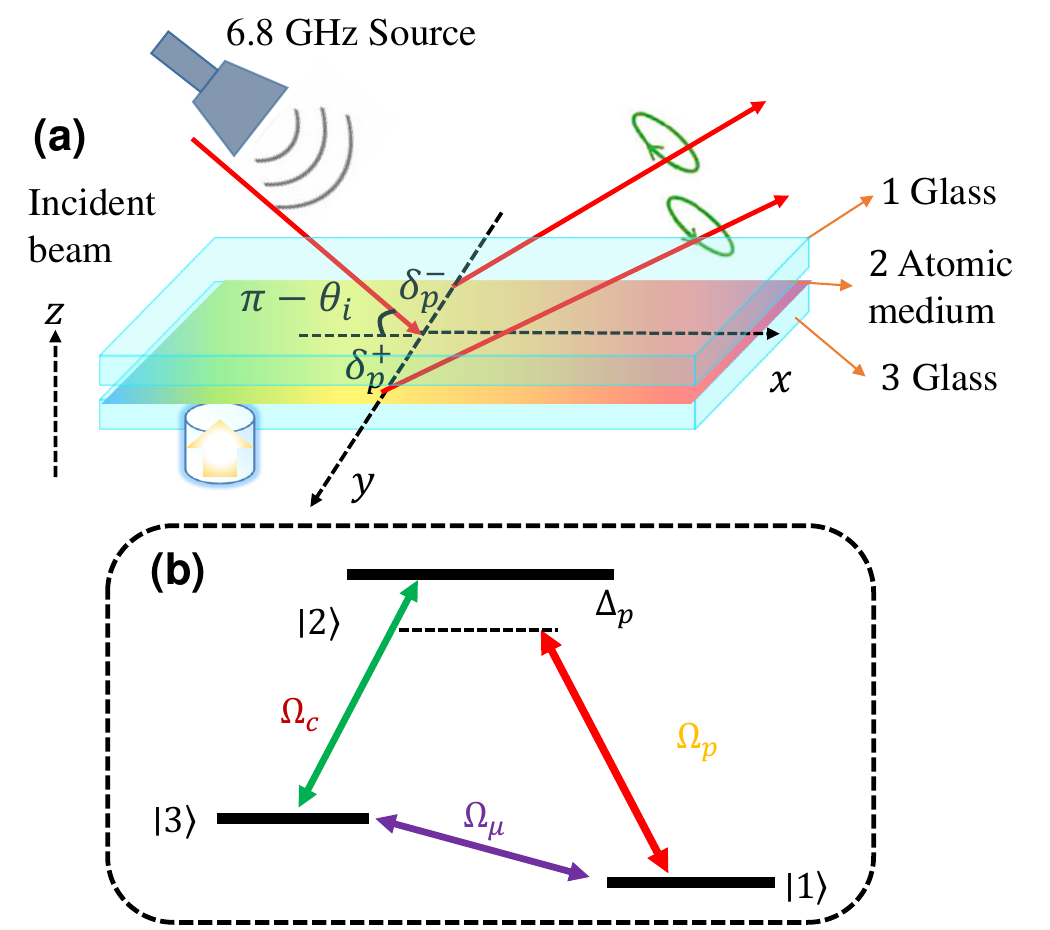}
\caption{(a) Schematic diagram of the prototype physical model. An ultracold atomic sample (shown by a rainbow sheet) spreads into an ultra-high vacuum glass chamber from the bottom nozzle. 
The TE and TM-polarized probe light beam is incident on the upper layer of the glass cell at an incident angle $\theta_{i}$. 
As a result, spin-dependent transverse displacement occurs for right circular $\delta_p^{+}$ and left-circular $\delta_p^{-}$ polarization components.
(b) Energy level diagram of the closed loop $\Lambda$-type atomic sample.}
\label{fig1}
\end{figure}

\section{Theory and Model}
The proposed prototype physical system is shown in Fig.~\ref{fig1} (a). We consider an ultracold dilute gas of $^{87}$Rb atoms contained within an ultra-high vacuum Cell or cavity
\footnote{Such micro cells, made of quartz or Pyrex with vapor thickness in the micrometers range, are available from Akatsuki Tech Japan. A typical refractive index for Pyrex glass is approximately 1.5, corresponding to $\epsilon \approx 2.25$.}
Such cold atoms can be transported from a standard magneto-optical trap using differential tubes or magnetic transport~\cite{greinerPhysRevA.63.031401}. Furthermore, the ultracold limit also eliminates the Doppler shift of the laser fields. 
The physical model is effectively a three-layer system. Layers 1 and 3 represent the upper and lower surfaces of the glass cell; each has thickness $d_1$ and permittivity $\epsilon_1$, while layer 2 is a trapped three-level atomic medium with thickness $d_2$ and permittivity $\epsilon_2$. 
The glass surfaces are assumed transparent, with outer surfaces having anti-reflection coating.
The energy level diagram of the atomic medium is shown in Fig.~\ref{fig1} (b).
The permittivity of the atomic medium is defined in terms of its dielectric susceptibility $\chi$ as $\epsilon_{2} = 1 + \chi$~\cite{wang_control_2008}. 
Here, $\chi=\chi^{\prime} +i \chi^{\prime \prime}$ is a complex quantity, where the real part $\chi^{\prime}$ represents dispersion and imaginary part $\chi^{\prime \prime}$ represents the absorption of the probe field as it interacts with the atoms.
The susceptibility is related to the refractive index of the atomic medium as $\eta = \sqrt{1 + \chi}$.

To determine $\chi$, we consider a $\Lambda$-type atomic system having an excited state $|2\rangle$ and two ground states $|1\rangle $ and $|3\rangle$ as shown in Fig.\ref{fig1}(b).
The optical probe field $E_{p}=\varepsilon_{p}e^{-i(\omega_{p}t-k_{p}z-\varphi_{p})}$ couples to ground state $|1\rangle$ and excited state $|2\rangle$. Here, $\omega_{p}$, $k_{p}$, and $\varphi_{p}$ are the frequency, wave number, and phase of the probe field, respectively. 
The optical control field $E_{c}=\varepsilon_{c}e^{-i(\omega_{c}t-k_{c}z -\varphi_{c})}$ couples energy state $|2\rangle$ and $|3\rangle$. 
For control field, frequency is $\omega_{c}$, wave number is $k_{c}$, and phase is $\varphi_{c}$.
To make the closed loop structure, the microwave field $E_{\mu}=\varepsilon_{\mu}e^{-i(\omega_{\mu}t-k_{\mu}z-\phi_{\mu})}$ of frequency $\omega_{\mu}$, wave number $k_{\mu}$, and an associated phase $\varphi_{\mu}$, couples the two lower energy states $|1\rangle$ and $|3\rangle$. 
The total Hamiltonian of the system is $\hat{H}=\hat{H}_{0}+\hat{H}_{int}$. 
The unperturbed part of the Hamiltonian is
\begin{eqnarray}
\hat{H}_{0} &=& \hbar(\omega_{1} \ket{1} \bra{1} + \omega_{2} \ket{2} \bra{2} + \omega_{3} \ket{3} \bra{3}),
\end{eqnarray} 
where $\omega_{1}$, $\omega_{2}$, and $\omega_{3}$ are the angular frequencies of respective atomic states. 
The interaction Hamiltonian of the system is written as
\begin{eqnarray}
 \hat{H}_{int}&=-\hbar[\Omega_{p}e^{-i\omega_{p}t}|2\rangle\langle 1|\nonumber +\Omega_{c}e^{-i\omega_{c}t}| 2\rangle\langle 3| \nonumber \\&+\Omega_{\mu}e^{-i\omega_{\mu}t}| 3\rangle\langle 1|+H.c.].
\end{eqnarray} 
The probe, control, and microwave fields have associated Rabi frequencies $\Omega_{p}=\wp_{12}E_{p}/\hbar$, $\Omega_{c}=\wp_{32}E_{c}/\hbar$, and $\Omega_{\mu}=\wp_{13}E_{\mu}/\hbar$, respectively. 
Here, $\wp_{ij}$ ($i,j=1,2,3$) is the dipole moment between respective transitions. 
Furthermore, Rabi frequencies are complex, denoted by $\Omega_{p,c,\mu}=|\Omega_{p,c,\mu}|e^{i\phi_{p,c,\mu}}$.

We apply the density matrix approach~\cite{scully1997quantum,scullyPRA, Saddique_2021} to obtain the matrix element $\rho_{21}$ for the transition $\ket{1} \rightarrow \ket{2}$.
We next consider slowly varying amplitudes $\rho_{21}= \tilde\rho_{21}\,e^{-i(\omega_p t-\phi_p)}$, $\rho_{23}= \tilde\rho_{23}\,e^{-i(\omega_c t-\phi_c)}$, and $\rho_{31}= \tilde\rho_{31}\,e^{-i[(\omega_p-\omega_c)t-(\phi_p-\phi_c)]}$, with the closed-loop resonance condition $\omega_p-\omega_c-\omega_\mu=0$.
This closed-loop condition is satisfied in the closed $\Lambda$-type system by taking $\Delta_p=\Delta_\mu$ and $\Delta_c=0$.
In this rotating frame of the probe field, the optical-Bloch equations for the three coherences become
\begin{subequations}
	\begin{align}
	\label{sigma21}
    \dot{\tilde\rho}_{21} & = -(\gamma_{21}-i\Delta_p)\tilde\rho_{21} -i|\Omega_p|(\rho_{22}-\rho_{11}) \nonumber\\& - i|\Omega_\mu|e^{-i\phi}\tilde\rho_{23} +i |\Omega_c|\tilde\rho_{31},\\
	\label{sigma23}
    \dot{\tilde\rho}_{23} & = -(\gamma_{23}-i\Delta_c)\tilde\rho_{23} -i |\Omega_c|(\rho_{22}-\rho_{33}) \nonumber\\&  + i |\Omega_p|\tilde\rho_{13} -i |\Omega_\mu|e^{+i\phi}\tilde\rho_{21}\\
	\label{sigma31}
    \dot{\tilde\rho}_{31}& = -(\gamma_{31}-i\Delta_\mu)\tilde\rho_{31} -i |\Omega_\mu|e^{-i\phi}(\rho_{33} -\rho_{11}) \nonumber\\ & -i |\Omega_p|\tilde\rho_{32} +i |\Omega_c|\tilde\rho_{21}.
	\end{align}
\end{subequations}
Here, $\phi = \varphi_{p}-\varphi_{c}-\varphi_{\mu}$ is the relative phase.
We define the small detunings as $\Delta_{p} = \omega_{p} - \omega_{21}$, $\Delta_{\mu} = \omega_{\mu} - \omega_{31}$, and $\Delta_{c} = \omega_{c} - \omega_{23}=0$, where $\omega_{21}$, $\omega_{31}$, and $\omega_{23}$ are the corresponding transition frequencies, respectively. 
The spontaneous decay rates are $\gamma_{21}$ for $|2\rangle \rightarrow |1\rangle$ transition and $\gamma_{31}$ for $|3\rangle \rightarrow |1\rangle$ transition.
We solved the coupled sets of equations (Eq.~\ref{sigma21} to Eq.~\ref{sigma31}) under steady state conditions, assuming that the control field is much stronger than the probe field $|\Omega_{c}| >> |\Omega_{p}|$, so that almost all the population remains in the ground state $\ket{1}$ ($\rho_{11}\approx 1,~\rho_{22}\approx 0,~\rho_{33}\approx 0$). As a result, one can obtain~\cite{scully1997quantum,scullyPRA}
\begin{eqnarray}\label{equ:rho}
\rho_{21}= \frac{{i(\gamma_{31}-i\Delta_{\mu})|\Omega_{p}|} - |\Omega_{c}||\Omega_{\mu}| e^{-i\phi}}{(\gamma_{31}-i\Delta_{\mu}) (\gamma_{21}-i\Delta_{p}) + |\Omega_{c}| ^{2}} .
\end{eqnarray}
The susceptibility $\chi=N|\wp_{12}|^2 \rho_{21}/\epsilon_{0}\hbar\Omega_{p}$, which determines the optical response of the probe field \cite{scully1997quantum}. Here, $N$ is the number of atoms per unit volume.

We consider the incident TE and TM-polarized probe light beam on the upper surface of the glass cell at incident angle $\theta_{i}$ as shown in Fig.~\ref{fig1}(a). 
This monochromatic Gaussian probe beam will be reflected at the structure interface or pass through the structure. In the reflection geometry, for a TM polarized Gaussian beam reflected by the interface, the field amplitudes of two circular components of reflected light can be expressed as~\cite{tan_enhancing_2016}:
\begin{equation}\label{a11}
\begin{aligned}
E_r^{ \pm} \propto & \frac{w_0}{w} \exp \left[-\frac{x_r^2+y_r^2}{w}\right] \\
& \times\left[r_p-\frac{2 i x_r}{k_0 w} \frac{\partial r_p}{\partial \theta_{i}} \mp \frac{2 y_r \cot \theta_{i}}{k_0 w}\left(r_p+r_s\right)\right].
\end{aligned}
\end{equation}
Here, $w=w_0\left[1+\left(2 \Lambda_{r} / k_0 w_0^2\right)^2\right]^{1 / 2}$ with beam waist $w_0$ and Rayleigh range $\Lambda_{r}=\pi w_0^2 / \lambda$. 
Here, $k_0=2 \pi / \lambda$ denotes the incident wave vector with $\lambda$ being the light wavelength.
The reflected light coordinate system is $\left(x_r, y_r, z_r\right)$, where superscript $\pm$ denotes left-hand circularly polarized (LHCP) and right circularly polarized (RHCP) photon states. 
In Eq.~\ref{a11}, the first-order Taylor series expansion of the Fresnel reflection coefficients is described as
\[
r_{p,s}(\theta_i+\delta\theta) 
\simeq
r_{p,s}(\theta_i) + \left.\frac{\partial r_{p,s}}{\partial\theta}\right|_{\theta_i}\delta\theta_i
\]
with respect to the small angular deviation $\delta\theta_i \simeq \delta k_x/k_1\cos\theta_i$.
In Eq.~\ref{a11}, the second term appears due to the Taylor-expansion contribution coming from the angle dependence of $r_p$. In contrast, the last term is the spin-orbit or geometric-polarization contribution responsible for photonic SHE.

Using the Method of Multiple Reflections~\cite{yariv_photonics_2007}, the complex reflection coefficients for TM-polarized $r_p$ and TE-polarized $r_s$ in glass-atom-glass structure can be written as \cite{waseem_gain-assisted_2024,wan_controlling_2020, salasnich2012enhancement,asiri_controlling_2023}
\begin{equation}\label{a33}
r_{p, s}=\frac{r_{p, s}^{12}+r_{p, s}^{23} e^{2 i k_{2 z} d_2}}{1+r_{p, s}^{12} r_{p, s}^{23} e^{2 i k_{2 z} d_2}},
\end{equation}
where $r_{p, s}^{i j}$ is the Fresnel's reflection coefficient at the $i$-$j$ interface (here $i,j=1,2,3$ for each layer).
The outer air-glass surfaces of the physical cell are not explicitly included in Eq.~\ref{a33} because the glass is transparent, and the outer cell surfaces may be anti-reflection coated in an experimental implementation. 
The upper glass directs the light to the atomic surfaces, and reflection happens at the glass-atom boundary. This scenario is analogous to the Kretschmann configuration in a surface plasmonics system~\cite{wan_controlling_2020}.
Therefore, Eq.~\ref{a33} describes the reflection from the two internal interfaces of this effective three-layer system, namely the glass-atom and atom-glass interfaces.
The complex reflection coefficients can also be calculated using an equivalent transfer matrix approach, which also gives the same physical results~\cite{yariv_photonics_2007, Born1999}.\\
For TM polarized, the Fresnel's reflection coefficient at the $i$-$j$ interface are given as
\begin{equation}\label{a44}
r_p^{i j}=\frac{k_{i z} / \varepsilon_i-k_{j z} / \varepsilon_j}{k_{i z} / \varepsilon_i+k_{j z} / \varepsilon_j},
\end{equation}
and TE polarized
\begin{equation}\label{a55}
r_s^{i j}=\frac{k_{i z}-k_{j z}}{k_{i z}+k_{j z}}.
\end{equation}
Here $k_{i z}=\sqrt{k_0^2 \varepsilon_i-k_x^2}$ represents the normal wave vector in the corresponding layer, and $k_x=\sqrt{\varepsilon_1} k_0 \sin \theta_{i}$ is the wave vector along the $x$ direction.

The transverse displacements can be computed as~\cite{tan_enhancing_2016}
\begin{equation}\label{a22}
\delta^{ \pm}_{p}=\frac{\iint y_r\left|E_r^{ \pm}\left(x_r, y_r, z_r\right)\right|^2 \mathrm{~d} x_r \mathrm{~d} y_r}{\int\left|E_r^{ \pm}\left(x_r, y_r, z_r\right)\right|^2 \mathrm{~d} x_r \mathrm{~d} y_r}.
\end{equation}
Utilizing the reflected field amplitude given by Eq.~\ref{a11}, the corresponding transverse spin-displacements $\delta_p^{+}$ and $\delta_p^{-}$ can be expressed in terms of the reflective coefficients for the three-layer system~\cite{xiang2017enhanced,waseem_gain-assisted_2024,chen_wide-angle_2022,zhou_controlling_2019}:
\begin{equation}\label{eq:shift}
\delta_p^{ \pm}=\mp \frac{k_1 w_0^2 \operatorname{Re}\left [1+\frac{r_s}{r_p} \right]  \cot \theta_{i}}{k_1^2 w_0^2+\left|\frac{\partial \ln r_p}{\partial \theta_{i}}\right|^2+\left|\left(1+\frac{r_s}{r_p}\right) \cot \theta_{i}\right|^2},
\end{equation}
with $k_{1}=\sqrt{\varepsilon_1} k_0$, $r_{s,p}=|r_{s,p}| e^{i \Psi_{s,p}}$, and $\Psi_{s,p}$ are the phases of Fresnel reflection coefficients $r_{s,p}$. 
The Eq.~\ref{eq:shift} is the photonic SHE formula under the first-order Taylor series expansion of the Fresnel reflection coefficients for which the angular spread is of the order of the small parameter $\delta \theta_i \sim 1/k_1 w_0$. 
Beam divergence $1/k_1 w_0$ must be sufficiently small compared with the angular scale on which the reflection coefficient $r_p(\theta)$ varies.
However, at the exact Brewster angle ($r_p=0$), the first-order approximation becomes fragile because the term $\left|\partial_\theta\ln r_p\right|^2$ can blow up.
Therefore, the first-order approximation is valid near, but not exactly at, the Brewster singularity.
This means that Eq.~\ref{eq:shift} is numerically unstable near Brewster incidence angle because it evaluates the shift in a ratio $r_s/r_p$ and logarithmic term $\partial_\theta\ln r_p$ that is ill-conditioned when \(r_p\) is very small.
The numerical stability method is discussed in Appendix~\ref{apend:stab} to avoid artificial divergences or spurious sign flips in our results.

More explicitly, Eq.~\ref{equ:rho} shows that the term proportional to $\Omega_c \Omega_\mu e^{-i\phi}$ introduces both the microwave Rabi frequency $\Omega_\mu$ and the relative phase $\phi$ into the atomic coherence $\rho_{21}$.
Since the probe susceptibility is given by $\chi \propto \rho_{21}/\Omega_p$, the microwave field modifies the effective permittivity $\epsilon_2=1+\chi$.
This effective permittivity changes the Fresnel reflection coefficients $r_s$ and $r_{p}$ in Eqs.~\ref{a33} to \ref{a55}.
Because the transverse spin-dependent shift in Eq.~\ref{eq:shift} depends explicitly on $r_p$, $r_s$, their ratio, and their angular variation, both the magnitude $\Omega_\mu$ and the relative phase $\phi$ of the microwave field can control the photonic SHE.
It may be mentioned that in the case of a dielectric medium, the Brewster angle condition ($r_p \to 0$) exists for TM ($p$) polarization at a certain incidence angle. In contrast, no such condition exists for TE ($s$) polarization. 
Therefore, TM ($p$-polarized) photonic SHE is significantly high due to the large ratio $|r_s/r_p|$ (as shown in Fig~\ref{fig:ratio}). 
Therefore, we only focus on TM ($p$-polarized) photonic SHE in our work.

\begin{section}{Results and discussion}
In this section, we present the results of our numerical simulations in detail.
We only present the transverse shift of the right circularly polarized photon spin-dependent component $\delta_{p}^{+}$ because the beam shifts for the two circular components are equal in magnitude and opposite in direction.
Here, we have selected the atomic parameters $|\Omega_{c}|=3 \gamma$, $|\Omega_{p}|=0.1\gamma$, $\gamma_{21}=5\gamma$, $\gamma_{31}=0.001\gamma$, and $\lambda=780$ nm from Ref \cite{scullyPRA}. 
Here, we define the numerical value of $\gamma=2 \pi \times 10^6$ Hz. For simplicity, we define all parameters in units of $\gamma$ (or MHz frequency).
Furthermore, we consider the situation such that $\omega_{p}-\omega_{c}-\omega_{\mu}=0$. This condition satisfies in closed $\Lambda-$type system by using $\Delta_{p}=\Delta_{\mu}$ and $\Delta_{c}=0$ \cite{scullyPRA}. 
The other parameters are $d_2 =0.4~\mu$m, $\epsilon_1=2.25$, and $w_0=50 \lambda$.

\begin{figure}[h]
	\centering
	\includegraphics[width=0.98\linewidth]{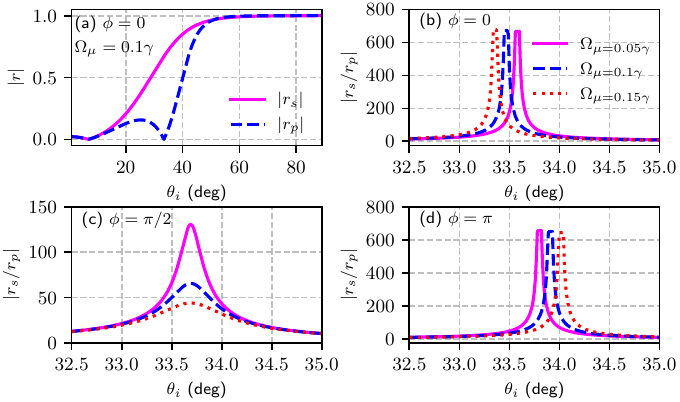}\\
\caption{(a) Individual reflection coefficient $|r_s|$ and $|r_p|$ verses incident angle $\theta_i$. 
The ratio $|r_{s}/r_{p}|$ as a function of $\theta_i$ at three fixed values of $\Omega_{\mu}=0.05 \gamma$ (solid curves), $\Omega_{\mu}=0.10 \gamma$ (dashed curves), and $\Omega_{\mu}=0.15 \gamma$ (dotted curves) (a) for $\phi=0$, (c) for $\phi = \pi/2$, and (e) for $\phi=\pi$. Here, $\Delta_{p}=0$.
}
\label{fig:ratio}
\end{figure}
\begin{figure*}[ht!]
	\centering
	\includegraphics[width=0.9\linewidth]{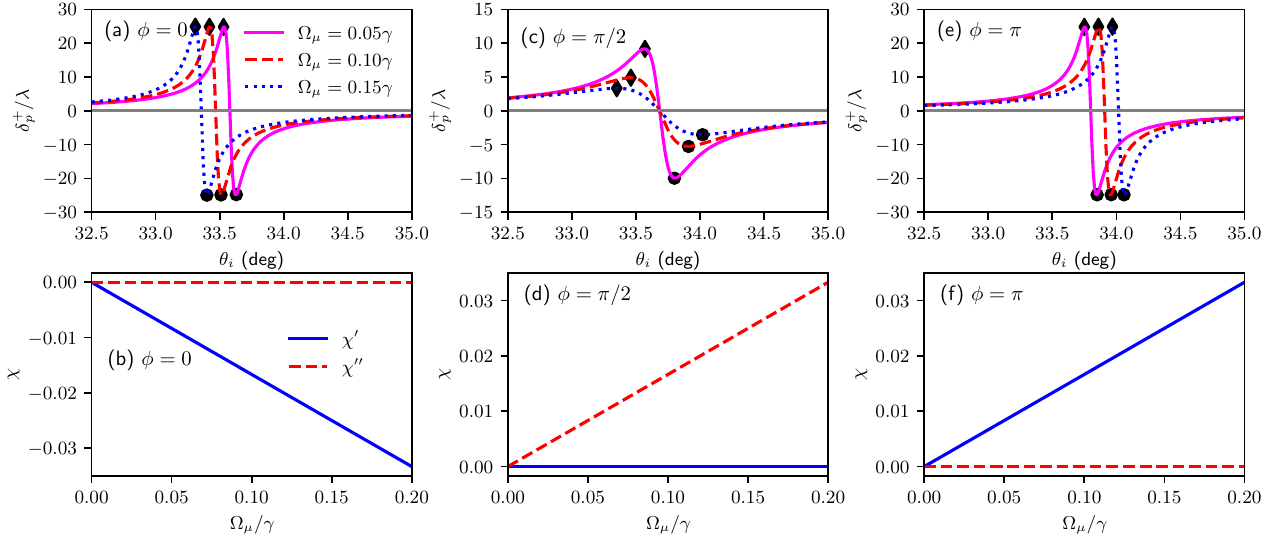}\\
\caption{Photonic SHE as a function of incident angle $\theta_i$ of the probe field at three fixed values of $\Omega_{\mu}=0.05 \gamma$ (solid curves), $\Omega_{\mu}=0.10 \gamma$ (dashed curves), and $\Omega_{\mu}=0.15 \gamma$ (dotted curves) is shown in (a) for $\phi=0$, (c) for $\phi = \pi/2$, and (e) for $\phi=\pi$. 
Diamonds and circles show the maximum and minimum value of the photonic SHE, respectively.
The corresponding real part of susceptibility $\chi^{\prime}$ (solid line) and imaginary part $\chi^{\prime \prime}$ (dashed line) are given in (b), (d), and (f). Here, we consider the resonant condition $\Delta_p = 0$.
}
\label{fig2:zero}
\end{figure*}
Photonic SHE explicitly depends on individual reflection coefficients $r_s$, $r_p$, and their ratio as evident from Eq.~\ref{eq:shift}. Therefore, we first plot reflection coefficients $r_s$ (solid curve) and $r_p$ (dashed curve) as a function of incident angle $\theta_i$ for the resonant condition $\Delta_p = 0$, $\Omega_{\mu}=0.1 \gamma$ and relative phase $\phi=\pi$, as shown in Fig.~\ref{fig:ratio} (a). 
Near the incident angle of $33.5^\circ$, $r_{p}$ approaches zero, while $r_s$ remains finite. The incident angle at which the reflection coefficient for $p$-polarized light vanishes ($r_p \rightarrow 0$) is known as the Brewster angle \cite{PRA_brewster}. 
The ratio of reflection coefficients $|r_s/r_p|$ is shown in Fig.~\ref{fig:ratio} (b) for $\phi=0$, (c) for $\phi=\pi/2$, and (d) for $\phi=\pi$. 
In each panel, we consider $\Omega_{\mu}=0.05 \gamma$ (solid curves), $\Omega_{\mu}=0.10 \gamma$ (dashed curves), and $\Omega_{\mu}=0.15 \gamma$ (dotted curves). Brewster angles position changes and the ratio $|r_{s}/r_{p}|$ significantly enhances at $r_{p}\approx0$, whose magnitude remains constant for each value of $\Omega_{\mu}$ at $\phi=0$ (Fig.~\ref{fig:ratio} (b)) and $\phi=\pi$ (Fig.~\ref{fig:ratio} (d)). 
For the case of $\phi=\pi/2$ in Fig.~\ref{fig:ratio}(c), the ratio $|r_{s}/r_{p}|$ decreases with the increase of $\Omega_{\mu}$ while Brewster angle position remains unchanged.

Next, we explore the influence of individual reflection coefficients and their ratio on photonic SHE.
The influence of the microwave field on the photonic SHE in a closed-loop $\Lambda$-type atomic system is illustrated in Fig.~\ref{fig2:zero} for the resonant condition $\Delta_p = 0$. The photonic SHE ($\delta_p^+/\lambda$) is plotted as a function of the incident angle $\theta_i$ for the relative phase $\phi=0$ in Fig.~\ref{fig2:zero}(a). The plots correspond to three fixed values of $\Omega_{\mu}=0.05 \gamma$ (solid curves), $\Omega_{\mu}=0.10 \gamma$ (dashed curves), and $\Omega_{\mu}=0.15 \gamma$ (dotted curves).
Here, we choose the range of incident angles around the point at which $r_p \rightarrow 0$ while $r_s$ is maximum from Fig.~\ref{fig:ratio}. 
These results in Fig.~\ref{fig2:zero}(a) show a significant enhancement of photonic SHE at the incident angle for which $|r_s/r_p|$ is large.
Near the Brewster angle ($r_{p} \rightarrow0$), variations in the Fresnel reflection coefficients lead to a pronounced change in the photonic SHE. 
Specifically, the photonic SHE initially increases to a positive maximum $(\delta_p^+/\lambda)_{\max}$ (shown by diamonds), then drops to zero precisely at the Brewster angle ($r_p=0$). Beyond this point, it rises again to reach a large negative value $(\delta_p^+/\lambda)_{\min}$ (shown by circles). This sign reversal occurs because the phase difference $\Psi_{s}-\Psi_p$ associated with reflection coefficients $r_s$ and $r_p$ undergoes a $\pi$-shift as the incident angle passes through the Brewster angle~\cite{waseem_gain-assisted_2024}. 
In this phase configuration, the Brewster angle, the angle $\theta_{i}^{\max}$ corresponding to $(\delta_p^+/\lambda)_{\max}$, and $\theta_i^{\min}$ corresponding to $(\delta_p^+/\lambda)_{\min}$, all changes linearly with increasing $\Omega_{\mu}$. 
This behavior is closely related to the ratio $|r_s /r_p|$ as seen in Fig.~\ref{fig:ratio}(b) and change in the medium's dielectric constant $\epsilon_2$, or its susceptibility $\chi$~\cite{wan_controlling_2020}. 
To illustrate this, we plot the real and imaginary parts of $\chi$ as a function of $\Omega_{\mu}$ in Fig.~\ref{fig2:zero}(b). As shown, $\chi$ is purely real and negative, resulting in a refractive index $\eta \approx 1+\chi /2 < 1$. An increase in the negative value of $\chi^{\prime}$ with higher $\Omega_{\mu}$ leads to a shift of the Brewster angle, $\theta_i^{\max}$, and $\theta_i^{\min}$ toward lower incident angles.

Next, we consider the case where the relative phase is set to $\phi = \pi/2$, while all other parameters remain unchanged. Figure~\ref{fig2:zero}(c) presents the photonic SHE as a function of the incident angle $\theta_i$.
In this configuration, both $(\delta_p^+/\lambda)_{\max}$ and $(\delta_p^+/\lambda)_{\min}$ decrease as $\Omega_{\mu}$ increases while the Brewster angle remains largely unaffected, unlike the case $\phi=0$.
Furthermore, $(\delta_p^+/\lambda)_{\max}$ and $(\delta_p^+/\lambda)_{\min}$ decrease as $\Omega_{\mu}$ increases due to decrease of ratio $|r_s/ r_p|$ as seen in corresponding Fig.~\ref{fig:ratio}(c).
At resonance $\Delta_p=\Delta_\mu=0$, $\chi \propto i\left(\gamma_{31}\Omega_p+\Omega_c\Omega_\mu\right) / \Omega_p$ from Eq.~\ref{equ:rho}, which is purely imaginary and positive. 
In this scenario, $\chi'\approx0$ while $\chi''$ increases with the strength of the microwave field as shown in Fig.~\ref{fig2:zero}(d).
The increase of $ \chi'' $ rounds the Brewster angle minimum of the $p$-polarized reflection, keeps $|r_p|$ finite, and reduces the effective ratio $|r_s/r_p|$ as seen in Fig.~\ref{fig:ratio}(c).
As a consequence, the photonic SHE decreases with increasing $\Omega_\mu$ at $\phi=\pi/2$, while the Brewster angle itself changes little because $\chi'\approx 0$.

We now consider the case $\phi = \pi$, as shown in Fig.~\ref{fig2:zero}(e) and (f). In this scenario, $\chi$ becomes purely real and positive ($\eta >1$), and increases with the strength of the microwave field $\Omega_{\mu}$. As a result, the magnitude of the photonic SHE remains stable to $w_0/2$, similar to the case with $\phi=0$. However, as $\chi'$ increases with $\Omega_{\mu}$, the Brewster angle, $\theta_i^{\max}$, and $\theta_i^{\min}$ shift toward higher incident angles. 
This behavior again closely resembles the variation of the ratio $|r_s /r_p|$ as seen in Fig.~\ref{fig:ratio}(d)).

The results shown in Fig.~\ref{fig2:zero} show that real $\chi^{\prime}$ ($\phi = 0, \pi$) causes the Brewster angle to shift with constant photonic SHE magnitude to the upper limit $w_0/2$, while imaginary $\chi^{\prime \prime}$ ($\phi = \pi/2$) reduces the SHE magnitude without affecting the Brewster angle. 
The larger ratio $|r_s/r_p|$ leads to larger photonic SHE $(\delta_p^{\pm}/\lambda)_{\max}$, which can be observed by making one-to-one correspondence between the results of ratio $|r_s/r_p|$ in Fig.~\ref{fig:ratio} and photonic SHE in Fig.~\ref{fig2:zero}.
It may be noted that the photonic SHE increases approximately linearly only while the denominator is still dominated by the constant term $k_1^2w_0^2+\left|\partial_{\theta_i}\ln r_p\right|^2$ in Eq.~\ref{eq:shift}.
After a finite maximum optimal ratio $|r_s/r_p|$, the photonic SHE decreases asymptotically as $\delta_p^\pm\sim 1/(r_s/r_p)$ for very large $|r_s/r_p|$. 
Furthermore, in the small beam divergence regime ($k_1w_0\gg 1$), inequalities $\left|\frac{\partial \ln r_p}{\partial\theta_i}\right|\ll k_1w_0$ and $\left| \left(1+\frac{r_s}{r_p}\right)\cot\theta_i \right|\ll k_1w_0$ holds, which is well satisfied in our case. 
Therefore, the first term dominates in Eq.~\ref{eq:shift}, yielding a simplified form $\delta_p^\pm \approx \mp (\lambda/2\pi) \left [ 1+\left|r_s/r_p\right|\cos(\Psi_s-\Psi_p) \right]\cot\theta_{i}$.
Hence, a larger ratio $\left|r_s/r_p\right|$ increases the photonic SHE only in the regime where the denominator remains nearly constant. 
Furthermore, the term $\cos(\Psi_s-\Psi_p)$ indicates the sign reversal between $\delta_p^+/\lambda$ and $\delta_p^-/\lambda$ across Brewster angle as seen in Fig.~\ref{fig2:zero} (a,c,e).

\begin{figure}[t]
\includegraphics[scale=0.9]{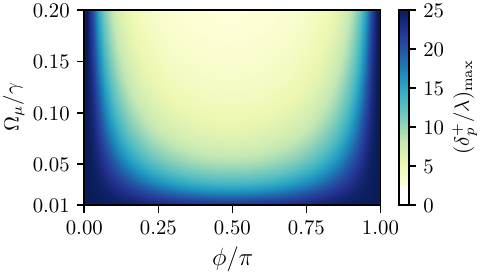}
\caption {(Color online) Density plot of maximum photonic SHE $(\delta_{p}^{+}/\lambda)_{\max}$ versus $\Omega_{\mu}$ and $\phi$ at $\Delta_p=0$. }
 \label{fig3} 
\end{figure}
In Fig.~\ref{fig3}, we present a density plot of $(\delta_p^+/\lambda)_{\max}$ as a function of the microwave field strength $\Omega_{\mu}$ and relative phase $\phi$ to illustrate these results further. 
The magnitude of the maximum photonic SHE remains constant versus $\Omega_{\mu}$ at $\phi = 0$ and $\phi = \pi$.
While for intermediate phase values, it stays maximum and constant at very low $\Omega_{\mu}$ but decreases rapidly as $\Omega_{\mu}$ increases. Faster decay of photonic SHE appears at $\phi=\pi/2$. Similarly, $(\delta_p^+/\lambda)_{\min}$ shows identical results with negative symmetry.
\begin{figure}[t]
\includegraphics[scale=1.0]{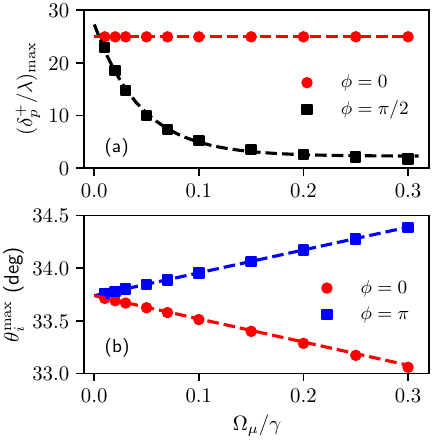}
\caption {(a) Maximum value of photonic SHE and (b) corresponding incident angle $\theta_i^{\max}$ versus amplitude of the microwave field $\Omega_{\mu}/ \gamma$ at $\Delta_p=0$.}
 \label{fig2:sensing} 
\end{figure}

The maximum value of photonic SHE $(\delta_p^+/\lambda)_{\max}$ for the two relative phases $0$ and $\pi$ stays constant at the value $w_0/2$ versus $\Omega_\mu/\gamma$ as shown in Fig.~\ref{fig2:sensing} (a) by solid circles.
The maximum value of photonic SHE equal to \(w_0/2\) is the intrinsic upper bound of the finite-Gaussian-beam centroid model rather than microwave-induced enhancement, attained when \(\left|\partial_{\theta_i}\ln r_p\right|\ll k_1 w_0\).
To justify this upper bound, let's define $x=\left(1+\frac{r_s}{r_p}\right)\cot\theta_i$ in Eq.~\ref{eq:shift}.
The susceptibility is predominantly real at \(\Delta_p=0\) and \(\phi=0,\pi\), and the extrema occur near Brewster-like conditions, so that \(\mathrm{Re}[x]\approx |x|\). Therefore, Eq.~\ref{eq:shift} approximated to $|\delta_p^\pm|\approx k_1 w_0^2\,x / (k_1^{2} w_0^2+x^2) \approx w_0/2$ when \(\left|\partial_{\theta_i}\ln r_p\right|\ll k_1 w_0\). 
Thus, the theoretical value $w_0/2$ is a beam-model saturation bound.
However, \(\chi''\neq 0\) for \(\phi=\pi/2\), then \(\varepsilon_2\) becomes complex. 
Consequently, $\mathrm{Re}[x] < |x|$, and the achievable maximum photonic SHE stays lower than \(w_0/2\) limit, which further decreases with the increase of absorption as shown by square data points in Fig.~\ref{fig2:sensing}(a).
Quantitatively, the decrease of $(\delta_p^+/\lambda)_{\max}$ follows an approximate exponential behavior with exponential constant of $24/\Omega_{\mu}$. This large exponential coefficient could be exploited for weak microwave-field amplitude sensing, i.e., microwave electrometry.

In Fig.~\ref{fig2:sensing}(b), quantitatively, the angular response for the two relative phases $0$ (circles) and $\pi$ (squares) is well approximated by linear trends as a function of $\Omega_\mu/\gamma$. 
A linear fit yields angular sensitivities $S_{\theta}^{(0)}\equiv d\theta_i/d(\Omega_\mu/\gamma)\approx-2.1$ ($\phi=0$) and $S_{\theta}^{(\pi)}\approx+2.25$ ($\phi=\pi$).
The two slopes have similar magnitudes but opposite signs. This enables a differential readout method that effectively doubles the sensitivity, allowing the smallest resolvable microwave change $\delta(\Omega_\mu/\gamma)_{\min} = \delta \theta_{\min}/|\Delta S|$ to be improved by a factor of two compared to the single-phase case. 
Therefore, the differential slope $\Delta S=S_\theta^{(\pi)}-S_\theta^{(0)}\approx4.27^\circ$ provides minimum detection limit of $\delta(\Omega_\mu/\gamma)_{\min}\approx2.3\times10^{-2}$ for an angular resolution of $0.1^\circ$.

\begin{figure}[t]
\includegraphics[scale=1.0]{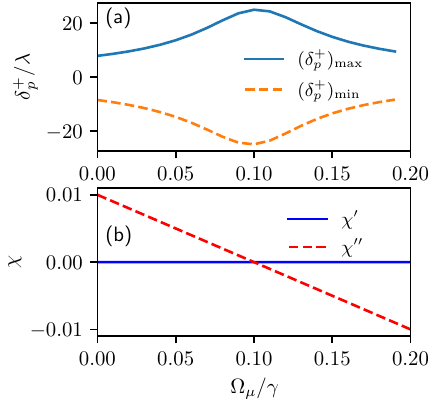}
\caption {(a) Maximum and minimum values of $(\delta_p^+/\lambda)$ as functions of $\Omega_{\mu}$ at $\phi = \pi$ and $\Delta_p = -\Omega_c$. (b) The corresponding plot of susceptibility $\chi$ as a function of $\Omega_{\mu}$. }
 \label{fig4:off} 
\end{figure}

When the microwave field $\Omega_{\mu}$ is set to zero, the response of the probe field exhibits standard EIT~\cite{harris1997electromagnetically, boller_observation_1991}, characterized by two absorption peaks at $\Delta_p = \pm \Omega_c$ and a transparency window (zero absorption) at $\Delta_p = 0$ (see Fig.~\ref{fig:chi}(a) in Appendix \ref{app1}).
Therefore, we now focus on the case $\Delta_p = \pm \Omega_c$. At these detunings, the absorption peaks can be either enhanced or suppressed, depending strongly on the relative phase $\phi$ between the optical fields and the strength of the microwave field $\Omega_{\mu}$. 
This occurs because applying a microwave field enables closed-loop transitions such as $\ket{1} \rightarrow \ket{2} \rightarrow \ket{3} \rightarrow \ket{1}$ and $\ket{1} \rightarrow \ket{3} \rightarrow \ket{2} \rightarrow \ket{1}$, where the coherence between the two ground states can interfere either constructively or destructively.
Consequently, the combination of optimized $\Omega_{\mu}$ and $\phi$ at $\Delta_p = \pm \Omega_c$ offers a promising mechanism for controlling the photonic SHE. 
Figure~\ref{fig4:off}(a) shows the maximum and minimum values of $(\delta_p^+/\lambda)$ as functions of $\Omega_{\mu}$ at $\phi = \pi$ and $\Delta_p = -\Omega_c$. 
The maximum and minimum value of photonic SHE increases initially and reaches its extremum at $\Omega_{\mu} = \Omega_p=0.1\gamma$, then decreases with further increase in $\Omega_{\mu}$. 
At this point, both the real and imaginary parts of the susceptibility $\chi$ vanish, resulting in $\epsilon_2 = 1$, i.e., a unit refractive index, as shown in Fig.~\ref{fig4:off}(b). This corresponds to an EIT‑like interference condition where both the dispersive and absorptive parts of the susceptibility vanish as explained in Appendix A. 
It is important to emphasize that the condition $\epsilon_2 = 1$ does not mean that the reflected probe beam vanishes. In the present glass–atom–glass structure, the reflected field is determined by the full three-layer Fresnel response, and therefore remains finite even when the atomic susceptibility becomes zero.
For $\Omega_{\mu} \neq \Omega_p$, $\chi$ becomes purely imaginary, which reduces the magnitude of the photonic SHE extrema similar to the case of $\Delta_p=0$ and $\phi=\pi/2$ as observed in Fig.~\ref{fig2:zero} (c). 
An identical behavior is observed for $\phi = 0$ and $\Delta_p = \Omega_c$, due to the symmetry of the system: changing the sign of $\Delta_p$ while flipping the phase by $\pi$ leads to the same interference conditions and susceptibility response (see Fig.~\ref{fig:chi} (b) $\&$ (c) in Appendix \ref{app1}).
As a result, the photonic SHE exhibits a similar feature.

\begin{figure}[t]
\includegraphics[scale=1.0]{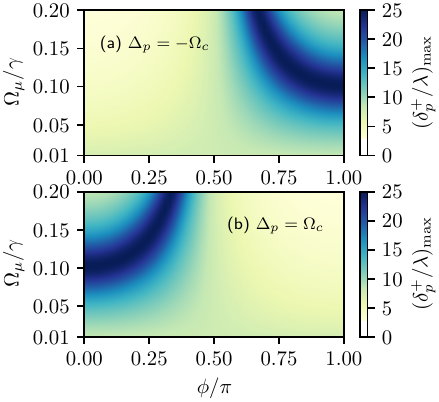}
\caption {(Color online) Density plot of maximum photonic SHE $(\delta_{p}^{+}/\lambda)_{\max}$ versus $\Omega_{\mu}$ and $\phi$ at (a) $\Delta_p=-\Omega_c$ and $\Delta_p=\Omega_c$}
 \label{fig5:off} 
\end{figure}

To examine the impact of relative phases other than $\phi = \pi$, we present a density plot of $(\delta_p^+/\lambda)_{\max}$ as a function of $\Omega_{\mu}$ and $\phi$ at $\Delta_p = -\Omega_c$ in Fig.~\ref{fig5:off}(a).
The intrinsic saturation value of the photonic SHE, $(\delta_p^+/\lambda)_{\max} = w_0 /2$ is reached at $\Omega_{\mu} = \Omega_p$ when $\phi = \pi$. 
Thus, the microwave field determines the parameter condition under which the bound is attained; it does not create or increase the bound itself.
For phases $\phi < \pi$, the same maximum value shifts to higher values of $\Omega_{\mu}$, i.e., $\Omega_{\mu} > \Omega_p$. 
Under this condition, the effective refractive index approaches unity due to the EIT-dominant response of the probe field.
When $\Omega_{\mu} < \Omega_p$, the maximum value of $(\delta_p^+/\lambda)$ remains below 15 for $\phi$ in the range $\pi/2$ to $\pi$, and drops further to below 8 when $\phi < \pi/2$.
A similar pattern is observed when the probe detuning is set to $\Delta_p = \Omega_c$ as seen in Fig.~\ref{fig5:off}(b), but mirrored along the phase axis. 
This mirrored behavior arises from the inherent symmetry of the closed-loop $\Lambda$-type system susceptibility (see Fig.~\ref{fig:chi} (b) $\&$ (c) in Appendix \ref{app1}): flipping the sign of the detuning (i.e., $\Delta_p \rightarrow -\Delta_p$) while also reversing the relative phase (i.e., $\phi \rightarrow -\phi$) does not change the underlying interference conditions between the optical and microwave pathways. 
As a result, the photonic SHE exhibits the same features, just reflected symmetrically across the $\phi=0$ axis.

\begin{figure}[ht!]
	\centering
	\includegraphics[width=0.85\linewidth]{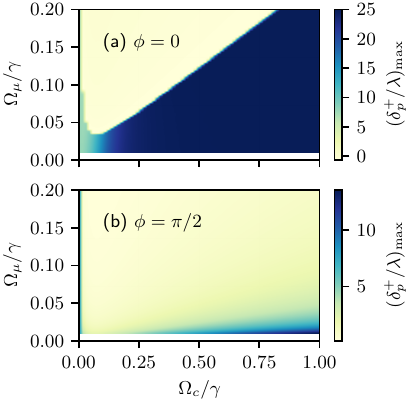}\\
\caption{Maximum value of photonic SHE $(\delta_p^+/\lambda)_{\max}$ as a function of control-field Rabi frequency $(\Omega_c )$ and the microwave-field Rabi frequency $(\Omega_\mu)$ at resonance $\Delta_p=0$.
}
\label{fig6:cont}
\end{figure}
In all previous results, the values of $\Omega_c=3 \gamma$ and $\Omega_p=0.1 \gamma$ were fixed under the condition of EIT, such that $\Omega_p \ll \Omega_c$~\cite{scullyPRA}. However, exploring the effects of the microwave field on the photonic SHE in the absence of EIT is also worthwhile.
Figure~\ref{fig6:cont} shows $(\delta_p^+/\lambda)_{\max}$ as a function of the control-field Rabi frequency $(\Omega_c / \gamma)$ and the microwave-field Rabi frequency $(\Omega_\mu / \gamma)$ at resonance, i.e., $\Delta_p=0$. Figure~\ref{fig6:cont}(a) corresponds to the relative phase $\phi=0$.
In the absence of the microwave field, $(\delta_p^+/\lambda)_{\max}$ increases rapidly in the low-control-field region from $\Omega_c=0$ to $\Omega_c=\Omega_p/2$, because the transparency window begins to open quickly. In the intermediate range, $\Omega_c=\Omega_p/2$ to $\Omega_c=2 \Omega_p$, $(\delta_p^+/\lambda)_{\max}$ grows slowly as the medium becomes increasingly transparent and EIT is almost fully developed. In the high-control-field region, beyond $\Omega_c=2 \Omega_p$, $(\delta_p^+/\lambda)_{\max}$ saturates at half the incident beam waist ($w_0/2$) due to vanishing absorption and dispersion. As a result, $(\delta_p^+/\lambda)_{\max}$ remains saturated for further increases in $\Omega_c$.
When the microwave field is applied above $\Omega_{\mu}=\Omega_p/2$, $(\delta_p^+/\lambda)_{\max}$ is strongly suppressed to zero in the low-control-field region. This occurs because the microwave field induces population redistribution and destructive interference between the dressed ground states. As the magnitude of the control field $\Omega_c$ increases, atomic coherence is re-established at a certain critical value of $\Omega_c$, and $(\delta_p^+/\lambda)_{\max}$ sharply jumps toward the saturation level $w_0/2$. This sudden jump could have potential applications in microwave-induced threshold switching of the photonic SHE.
The suppression of $(\delta_p^+/\lambda)_{\max}$ extends over a broader range of $\Omega_c / \gamma$ for stronger microwave coupling $\Omega_\mu$. Consequently, a larger control field $\Omega_c$ is required to restore EIT and recover $(\delta_p^+/\lambda)_{\max}$. These results indicate that the microwave field provides a flexible mechanism to modulate and switch the photonic SHE.
It may be noted that the behavior shown in Fig.~\ref{fig6:cont}(a) represents threshold switching rather than bistable or hysteretic switching.
We note that the results for the relative phases $\phi=\pi$ and $\phi=0$ are identical, which is consistent with the results shown in Fig.~\ref{fig3}.

In Fig.~\ref{fig6:cont}(b), we consider the case where the relative phase is $\phi=\pi/2$. Unlike the cases of $\phi=0$ or $\pi$, $(\delta_p^+/\lambda)_{\max}$ is suppressed over the entire range of $\Omega_c / \gamma$ when the microwave field is applied. In other words, sharp switching of $(\delta_p^+/\lambda)_{\max}$ is suppressed across the full range of $\Omega_c / \gamma$. Only with sufficiently large values of $\Omega_c$ and weak $\Omega_{\mu}$ does a slow recovery of $(\delta_p^+/\lambda)_{\max}$ appear. Hence, the phase of the microwave field acts as a tunable parameter that enables a continuous transition from controllable switching to complete suppression of $(\delta_p^+/\lambda)_{\max}$.
\end{section}

\section{Experimental feasibility}
In this section, we discuss the experimental feasibility of our proposal. 
A feasible experimental implementation can be constructed by combining the microwave-controlled closed $\Lambda$ atomic-coherence scheme of \cite{scullyPRA} with the weak-measurement photonic SHE detection protocol of~\cite{hosten_observation_2008}.
Such a $\Lambda$-type scheme can be realized experimentally by using the lowest ground-state hyperfine levels and coupling them to a single excited-state hyperfine level of $^{87}$Rb~\cite{scullyPRA}. 
In $^{87}$Rb, employing the $D_1$ transition line, the configuration is as follows: lowest ground state $|1\rangle =\ket{5S_{1/2},F=1}$, upper ground state $|3\rangle=\ket{5S_{1/2},F=2}$, and an excited state $|2\rangle =\ket{5P_{1/2},F^{'}=2}$. 
Here, $F$ or $F^{'}$ denotes the total atomic angular momentum quantum number in the hyperfine structure.
The $|1\rangle\leftrightarrow|3\rangle$ hyperfine transition can be driven by applying standard microwave or radio frequency spectrometry; a microwave field or radio frequency signal can be applied through an antenna \cite{nagase_measurement_2025}. Another feasible option is to use an electro-optical modulator to generate a sideband of frequency 6.8 GHz~\cite{scullyPRA}.
Optical frequencies for probe and control fields can be prepared using an acoustic optical modulator (AOM).

\begin{figure}[h]
	\centering
	\includegraphics[width=0.8\linewidth]{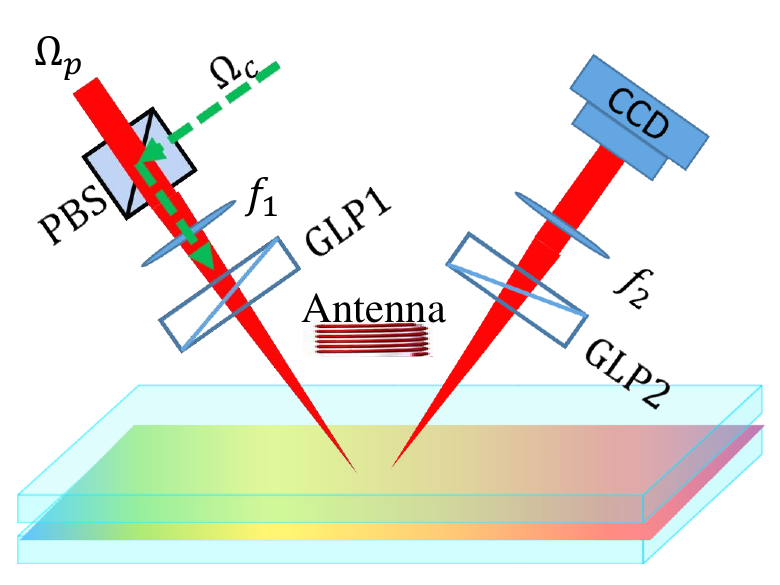}\\
\caption{Proposed schematic of the experimental setup to detect the photonic SHE in atomic vapors. A polarized Gaussian probe beam $\Omega_p$ is incident at an angle on the glass-atom-glass interface. Control field $\Omega_c$ is set to co-propagate with the probe field by combining at a polarization beam splitter (PBS).
The lenses with a focal length $f_1$ and $f_2$ can focus and collimate the probe beam.
The Glan laser polarizers (GLP1 and GLP2) select the preselected and postselected states, respectively. The CCD captures the intensity profile of the reflected beam. The radio frequency antenna drives the microwave field transition.
}
\label{fig7:exp}
\end{figure}
The proposed experimental scheme is shown in Fig.~\ref{fig7:exp}.
On the optical readout side, the experiment follows the quantum weak-measurement protocol employed for photonic SHE detection~\cite{liu_photonic_2022,hosten_observation_2008}.
A weak probe beam at $780~\mathrm{nm}$ first passes through a lens of focal length $f_1$, which sets the beam waist at the sample interface and thereby determines the Rayleigh range that enters the weak-measurement amplification factor.
A Glan–laser polarizer (GLP1) prepares the input probe beam in the desire pre-selection polarization state.
The probe beam is incident on the interface near the Brewster angle, where the reflected field acquires the spin-dependent transverse displacement predicted by the photonic SHE model.
After reflection from the glass-atom-glass structure, the beam enters the postselection stage, which consists of a Glan-laser polarizer (GLP2) with a small postselection offset angle $\alpha$. This postselection converts the small spin-dependent transverse displacement into an amplified centroid shift, which is then imaged by the second lens of focal length $f_{2}$ onto a CCD or position-sensitive detector. Control field $\Omega_c$ is set to co-propagate with the probe field by combining the two beams at a polarization beam splitter (PBS).

Based on this standard weak-measurement readout parameter~\cite{hosten_observation_2008} and our atomic model parameters~\cite{scullyPRA}, we performed the shot-noise (SN) limit analysis for the results shown in Fig.~\ref{fig2:sensing}.
Using a standard weak-measurement readout for $\lambda=780\,\mathrm{nm}$, $w_{0}=50~\mu\mathrm{m}$, input probe power before postselection $P_{\rm in}=1~\text{mW}$, detection efficiency $\eta=0.8$, $f_{1}=25~\mathrm{mm}$, $f_{2}=35~\mathrm{mm}$, and $\alpha=1^\circ$, we obtain a weak-measurement amplification factor $A_{\rm WM}=(f_{2}/\Lambda_r) \cot \alpha\approx199$ and corresponding intrinsic photonic SHE displacement sensitivity at the shot-noise limit of $\delta_{\rm SN}\approx1.12~\mathrm{pm}/\sqrt{\mathrm{Hz}}$~\cite{hosten_observation_2008}.
For exponential decrease of maximum photonic SHE value at $\phi=\pi/2$ in Fig.~\ref{fig2:sensing}(a), the minimum resolvable normalized microwave sensitivity is
$\delta\left(\Omega_\mu/\gamma\right)_{\mathrm{max}} =\delta_{\mathrm{SN}}/(\lambda\left| \partial(\delta_{p}^{+}/\lambda)_{\max}/ \partial(\Omega_\mu/\gamma) \right|)$.
Using the fitted peak response in Fig.~\ref{fig2:sensing}(a) and evaluating its slope at $\Omega_\mu/\gamma=0.05$, we estimated corresponding electric field ($\Omega_\mu = \wp_{13} E_\mu/\hbar$) sensitivity of $1.68~ \mu\mathrm{V/cm}/\sqrt{\mathrm{Hz}}$.
Similarly, for the differential angular results shown in Fig.~\ref{fig2:sensing}(b), the normalized microwave angular sensitivity is $\delta\left(\Omega_\mu / \gamma \right)_{\mathrm{diff}} = \delta_{\mathrm{SN}}/(\lambda\left| \partial(\delta_p^{+}/\lambda)/\partial\theta_i \right| |\Delta S|)$, with $\Delta S=S_\theta^{(\pi)}-S_\theta^{(0)} \simeq4.27^\circ$.
Using the results shown in Fig.~\ref{fig2:sensing}(b), we estimate a differential angular sensitivity of order $(1.20-1.45)~ \mu\mathrm{V/cm}/\sqrt{\mathrm{Hz}}$.
These estimated electric-field sensitivities are lower (improved sensitivity) than both the $\sim 3~\mu\mathrm{V/cm}/\sqrt{\mathrm{Hz}}$ estimated in Ref.~\cite{zhou_improving_2023} and the $8~\mu\mathrm{V/cm}/\sqrt{\mathrm{Hz}}$ reported in Ref.~\cite{sedlacek_microwave_2012} for Rydberg‑atom microwave electrometry schemes.

We neglect Doppler broadening because we consider an ultracold dilute $^{87}\mathrm{Rb}$ ensemble, for which the spread of atomic velocities is negligible.
However, for a thermal vapor in the present closed-loop microwave $\Lambda$-type system, the Doppler-shifted detunings are $\Delta_p(v)=\Delta_p-k_p v$, $\Delta_c(v)=\Delta_c-k_c v$, and $\Delta_\mu(v)=\Delta_\mu-k_\mu v $.
Then, the closed loop detuning becomes $\delta_{\rm loop}(v)=\Delta_p(v)-\Delta_c(v)-\Delta_\mu(v) = (\Delta_p-\Delta_c-\Delta_\mu) -(k_p-k_c-k_\mu)v$. 
Since the microwave wave number $k_\mu$ is much smaller than the optical wave numbers $k_\mu \ll k_p,\;k_c$, the microwave contribution to the Doppler shift is negligible.
The corresponding two-photon (Raman) detuning is therefore $\delta_{2\gamma}(v)=\Delta_p(v)-\Delta_c(v) = (\Delta_p-\Delta_c)-(k_p-k_c)v$.
As shown in Fig.~\ref{fig7:exp}, arranging the probe and control fields to co-propagate by combining them at the PBS and with comparable wavelengths of 780~nm, such that $k_p \approx k_c$, the two-photon detuning becomes nearly Doppler-free, $\delta_{2\gamma}(v)\approx \Delta_p-\Delta_c$, especially for $\Lambda$-level atomic structure~\cite{fleischhauer_electromagnetically_2005}.
Thus, although the one-photon optical resonances remain Doppler-broadened, the two-photon Raman resonance responsible for the EIT-based atomic coherence remains nearly velocity-independent.
Furthermore, dephasing is already included phenomenologically in the density-matrix model (Eq.~\ref{equ:rho}) through the coherence decay rates, which are fixed for the atomic structure shown in Fig.~\ref{fig1}(b).
Technical noise, such as detector noise and finite angular resolution, would increase the experimental uncertainty in determining the photonic SHE peak value and angular position.

\begin{section}{Conclusion}

In conclusion, we have explored the photonic SHE tunability and microwave electrometry in a closed-loop $\Lambda$-type atomic system by manipulating the strength of an applied microwave field and the relative phase between the driving fields. 
When the relative phase is set to $\phi = 0$ or $\phi = \pi$ at $\Delta_p=0$, the photonic SHE magnitude remains near the intrinsic finite-beam saturation limit $w_0/2$~\cite{kimhalfwaist}.
However, varying microwave field the positions of the Brewster angle, $(\delta_p^+/\lambda)_{\max}$, and $(\delta_p^+/\lambda)_{\min}$ shift either toward lower or higher incident angles depending on the sign of the susceptibility $\chi$, showing a differential angular sensitivity of order $(1.20-1.45)~ \mu\mathrm{V/cm}/\sqrt{\mathrm{Hz}}$.
For intermediate values of $\phi$, especially around $\pi/2$, the photonic SHE becomes exponentially tunable in magnitude, providing another way of microwave sensing in the weak electric field limit with estimated sensitivity of order $1.68~ \mu\mathrm{V/cm}/\sqrt{\mathrm{Hz}}$ at $\Omega_\mu/\gamma=0.05$.
Furthermore, at $\Delta_p = \pm \Omega_c$, intrinsic saturation limit of the photonic SHE up to $w_0 /2$ is achievable by setting $\Omega_{\mu} \geq \Omega_p$ and appropriate phase $\phi$.
Adjusting the microwave amplitude at a certain critical value of the control field shows a phase-control-switchable photonic SHE.
Our theoretical model can be extended to multi-level Rydberg systems~\cite{chopinaud_optimal_2021}, where much larger microwave dipole moments may enable enhanced precision of microwave electrometry based on photonic SHE.
Furthermore, our proposed scheme is experimentally feasible by combining well-established methods of atomic coherence manipulation~\cite{scullyPRA} and weak-value-based photonic SHE measurements~\cite{hosten_observation_2008}, thereby motivating future experimental investigations.

Furthermore, the total spin-dependent splitting in the photonic SHE is also complex and given by $\delta_t = \delta_{p}^{\pm} + i \delta_k^{\mp}$~\cite{zheng_measurement_2024, chen2021measurement, salasnich2012enhancement}
The real part, $\delta_{p}^{\pm}$, determines the spin-dependent spatial splitting in position space, which we discuss extensively in this work. The imaginary part, $\delta_{k}^{\mp}$, is associated with spin-dependent splitting in momentum space at propagation distance $z$, and manifests as an angular shift $\Theta^{\mp} = z\delta_k^{\mp} / \Lambda_r$~\cite{chen2021measurement}.
In the post-selection procedure often used for weak-value measurement (see Section IV), the measured spatial photonic SHE signal also contains a contribution from the angular photonic SHE. Spatial and angular shifts exhibit different sensitivities to various physical parameters.
Recently, a modified weak-value scheme was demonstrated using post-selection with a real weak value (by including quarter-wave plates)~\cite{zheng_measurement_2024}. In this scheme, the spatial SHE is suppressed and the angular SHE is selectively amplified for detection. It would be interesting in future work to treat the angular shift as the pointer for microwave sensing in a dedicated study.

\end{section}

\appendix
\section{Susceptibility analysis}\label{app1}

The atomic susceptibility $\chi=N|\wp_{12}|^2 \rho_{21}/\epsilon_{0}\hbar\Omega_{p}$ consists of real part $\chi'=\chi'_0-A\cos\phi-B\sin\phi$ and imaginary part  $\chi''=\chi''_0+A\sin\phi-B\cos\phi$. Defining proportionally constant $P=N|{\wp}_{12}|^2/(\varepsilon_0\hbar\,\Omega_p)$ and after simplifications in Eq.~\ref{equ:rho}, we obtained mathematical forms of coefficients
\[
A=P\,\frac{|\Omega_c||\Omega_\mu|\,X}{X^2+Y^2},
\qquad
B=P\,\frac{|\Omega_c||\Omega_\mu|\,Y}{X^2+Y^2},
\]
and offset terms
\[
\chi'_0=P\,\Omega_p\,\frac{\Delta_\mu X-\gamma_{31}Y}{X^2+Y^2},
\qquad
\chi''_0=P\,\Omega_p\,\frac{\Delta_\mu Y+\gamma_{31}X}{X^2+Y^2}.
\]
with 
\[
X=\gamma_{31}\gamma_{21}-\Delta_\mu\Delta_p+|\Omega_c|^2,
\qquad
Y=\gamma_{31}\Delta_p+\gamma_{21}\Delta_\mu.
\]
\setcounter{figure}{0}
\renewcommand{\thefigure}{A\arabic{figure}}
\begin{figure}[h]
	\centering
	\includegraphics[width=0.98\linewidth]{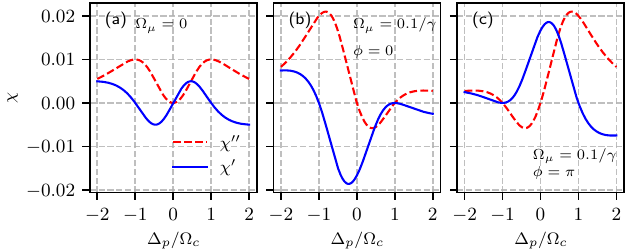}\\
\caption{Dispersion $\chi'$ (solid curves) and absorption $\chi''$ (dashed curves) versus $\Delta_p/\Omega_c$ for (a) $\Omega_{\mu}=0$, (b) $\Omega_{\mu}=0.1 \gamma, ~ \phi=0$, and (c) $\Omega_{\mu}=0.1 \gamma, ~ \phi=\pi$. 
}
\label{fig:chi}
\end{figure}
Fig.~\ref{fig:chi} shows dispersion $\chi'$ (solid curves) and absorption $\chi''$ (dashed curves) versus $\Delta_p/\Omega_c$.
In Fig.~\ref{fig:chi}(a), when $\Omega_{\mu}=0$, the response of the probe field exhibits a standard EIT, characterized by a transparency window at $\Delta_p=0$ and two absorption peaks at $\Delta_p=\pm \Omega_c$.  
Fig.~\ref{fig:chi}(b) and (c) represents the results for $\Omega_{\mu}=0.1 \gamma, ~ \phi=0$, and  $\Omega_{\mu}=0.1 \gamma, ~ \phi=\pi$, respectively.
At \(\phi=0\), the real part satisfies $\chi'(0)=\chi'_0-A$ while imaginary part $\chi''(0)=\chi''_0-B$.
At \(\phi=\pi\), the real part satisfies $\chi'(\pi)=\chi'_0+A$ while the imaginary parts $\chi''(\pi)=\chi''_0+B$.
Therefore, the two phases produce equal magnitude but opposite sign relative to the offset values, 
Thus, changing $\phi$ from 0 to $\pi$ produces equal magnitude but opposite sign relative to the offset values, leading to symmetric behavior in the photonic SHE response as clearly seen in Fig.~\ref{fig:chi}(b) and (c).  
Therefore, \(\phi=0\) and \(\phi=\pi\) are symmetry-paired points.
This is exactly the symmetry observed in Fig.~\ref{fig2:zero}(b), Fig.~\ref{fig2:zero}(f), and Fig.~\ref{fig5:off} in the main text.

\section{Numerical Stability}\label{apend:stab}
Eq.~\ref{eq:shift} is numerically unstable near Brewster incidence because it evaluates the shift in a ratio $r_s/r_p$ and logarithmic term $\dfrac{\partial \ln r_p}{\partial \theta_i}$ that is ill-conditioned when \(r_p\approx 0\).
Therefore, for numerical stability, we multiply the numerator and denominator by \(|r_p|^2\) in Eq.~\ref{eq:shift}, to obtain a stable equivalent form:
\begin{equation}
   \delta_p^\pm = \mp\, \frac{ k_1 w_0^{\,2}\, \mathrm{Re}\!\bigl[r_p^*(r_p+r_s)\bigr]\cot\theta_i }{ k_1^{2} w_0^{\,2}|r_p|^{2} + \left|\frac{\partial r_p}{\partial \theta_i}\right|^{2} + \left|(r_p+r_s)\cot\theta_i\right|^{2} }. 
\end{equation}
This stable form is expressed in terms of the complex coefficients \(r_p\), \(r_s\) rather than the ratio \(r_s/r_p\). The derivative \(\partial_\theta \ln r_p\) is evaluated using a symmetric finite-difference scheme, rather than by differentiating \(\ln r_p\) to avoid numerical spikes. All the results presented in the main text are evaluated using this numerically stabilized form.

 \section*{acknowledgments}

MW would like to acknowledge the fruitful discussion with Dr. Shahid Qamar and Dr. Muhammad Irfan. 
  
 \section*{Conflict of interest}
 The authors declare that they have no conflict of interest.

 \section*{Data availability}
Data underlying the results presented in this paper may be obtained from the authors upon reasonable request and are also reproducible from theoretical models.

\bibliographystyle{apsrev4-2}
\bibliography{ref.bib}

\end{document}